\documentclass[aps,prl,superscriptaddress,preprintnumbers,
showpacs,legalpaper,twoside,twocolumn]{revtex4}

\usepackage{bm}
\usepackage{graphicx}
\usepackage{amsfonts}
\usepackage{amsmath}
\usepackage{amssymb}

\begin{document}

\title{Entanglement Entropy in the Two-Dimensional
Random Transverse Field Ising Model}

\author{Rong Yu}
\affiliation{Department of Physics and Astronomy, University of
Southern California, Los Angeles, CA 90089-0484}
\affiliation{Department of Physics and Astronomy, University of
Tennessee, Knoxville, TN 37996-1200}
\author{Hubert Saleur}
\affiliation{Service de Physique Th\'{e}orique, CEN Saclay, Gif Sur
Yvette, F-91191} \affiliation{Department of Physics and Astronomy,
University of Southern California, Los Angeles, CA 90089-0484}
\author{Stephan Haas}
\affiliation{Department of Physics and Astronomy, University of
Southern California, Los Angeles, CA 90089-0484}

\pacs{03.65.Ud, 75.10.Nr, 73.43.Nq, 64.60.Ak}

\begin{abstract}
The scaling behavior of the entanglement entropy in the
two-dimensional random transverse field Ising model is studied
numerically through
 the strong disordered renormalization group method. We
find  that the leading term of the entanglement entropy always
scales linearly
with the block size. However, besides this \emph{area law}
contribution, we find a subleading logarithmic correction at the
quantum critical point. This correction is
discussed from the point of view  of an underlying  percolation
transition, both at finite and at zero temperature.

\end{abstract}

\maketitle

The study of novel quantum phases and related quantum phase
transitions (QPTs) is at the forefront of many recent developments
in condensed matter physics. It relies heavily on the concept of
entanglement entropy.

A state $|\Phi\rangle$ of a bipartite system $A\bigcup B$ is
entangled if it cannot be described accurately in either subsystem
$A$ or $B$. A convenient measure of this entanglement is
  the  entropy, $S_A=-Tr\rho_A
\log_2\rho_A$, where $\rho_A=Tr_B |\Phi\rangle\langle\Phi|$. Denote
the linear dimensions of $A\bigcup B$ and $A$ as $M$ and $L$,
respectively. An important question in quantum many-body systems is
to study how $S_A(L)$ scales with $L$ in the limit of
$M\rightarrow\infty$ in different quantum phases. This question has
been extensively investigated in one dimension
(1D)~\cite{Vidaletal03, Latorreetal04, RefaelMoore04, Laflorencie05,
Skrovseth05, Crameretal06, CalabreseCardy04}. There, it is now well
understood that for non-critical systems, $S(L)$ saturates to a
constant as $L\rightarrow\infty$; whereas in critical systems, a
logarithmic modification stands out as the leading term: $S(L)\sim
\ln L$, and its coefficient is associated with the central charge of
the related $(1+1)$ conformal field theory
(CFT)~\cite{CalabreseCardy04}. In higher dimensions, it is generally
believed that an \emph{area law} holds at least for non-critical
systems: the entanglement entropy scales as the area of the boundary
between subsystem $A$ and $B$, $S(L)\sim
L^{d-1}$. 
This has been confirmed by studies on bosonic harmonic lattice
systems~\cite{Crameretal06, Plenioetal05}. For critical systems, the
situation is more complicated. The area law is shown to be violated
in free fermion systems with a finite Fermi surface~\cite{Fermi,
Li06}. But it still holds for fermionic systems without a finite
Fermi surface~\cite{Li06}, and critical bosonic
systems~\cite{BosonCr}.

In $d$-dimensional system where the area law holds, $S(L)\sim f_s
L^{d-1}$. Here $f_s$ is a boundary free energy determined by the
short-distance properties of the system, and is hence not universal.
It is thus interesting to wonder about subleading ~\cite{footnote}
terms in $S(L)$, where, maybe, universal coefficients depending only
on the model and the topological properties of the system could
appear. For instance, it was recently found that in two-dimensional
(2D) gapped systems, a subleading constant contribution in $S(L)$ is
related to the topological order~\cite{topo}. Also, for a class of
$z=2$ conformal quantum critical systems in 2D, a universal
logarithmic correction to the area law term has been
found~\cite{FradkinMoore06}. Clearly, the problem is not fully
settled.

It is of course also possible to investigate entanglement in quantum
disordered systems. In a series of studies in 1D based on strong
disorder renormalization group (SDRG) techniques~\cite{SDRGreview},
it was found that for the class of 1D infinite randomness fixed
points (IRFP), a $\ln L$ term in $S(L)$ is also
present~\cite{RefaelMoore04,random1d}. Similar results were also
discovered for 1D aperiodic systems~\cite{aperiod}.

\begin{figure}[h]
\begin{center}
\includegraphics[
bbllx=-40pt,bblly=180pt,bburx=550pt,bbury=430pt,%
     width=85mm,angle=0]{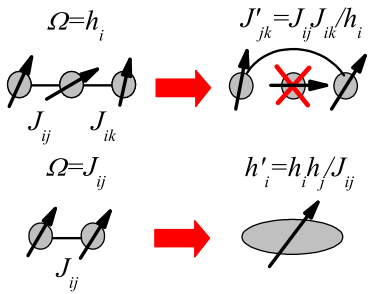}
\caption{(color online) Basic RG transformations (see text for
detail). Upper: energy scale is a field; lower: energy scale is an
Ising coupling.} \label{F.RG}
\end{center}
\end{figure}

In this paper, we report on our study of the
 2D random transverse
field Ising (RTFI) model, and the numerical calculation of the
entanglement entropy using the SDRG technique. The model is defined
on a 2D square lattice with linear dimension $M$ and open boundary
condition. The subsystem $A$ is a $L\times L$ square region located
in the center of the square lattice. The Hamiltonian reads
\begin{equation}\label{E.Hamiltonian}
H = -\sum_{\langle i,j\rangle} J_{ij} S^z_i S^z_j - \sum_i h_i
S^x_i.
\end{equation}
The Ising coupling $J_{ij}$ and the transverse field $h_i$ take
random values drawn from the following box  shape distributions:
\begin{eqnarray}\label{E.Distrib}
P(J)=\Theta(J)-\Theta(J-1),\nonumber\\
P(h)=\frac{1}{h_0} \left[\Theta(h)-\Theta(h-h_0)\right].
\end{eqnarray}
This model is known to have a quantum phase transition which is
governed by an IRFP~\cite{2DRandomIsing,Motrunich00}. Here the
critical point is tuned by $h_0$. Starting from the original
Hamiltonian Eq.~\ref{E.Hamiltonian}, the SDRG finds the ground state
by successively eliminating the highest energy degrees of
freedom~\cite{SDRGreview, Fisher94}. At each RG step, we look for
the largest term in the  Hamiltonian; its coupling (or field) is
defined as the energy scale $\Omega$ at this step. If $\Omega =
h_i$, the local spin is frozen in the eigenstate of $S^x_i$ by the
local field. It is then eliminated from the system, and an effective
coupling $J'_{jk}\thickapprox max(J_{jk},J_{ji}J_{ik}/h_i)$ is
introduced between its two neighboring spins at sites $j$ and $k$.
If $\Omega = J_{ij}$, the two spins involved respond to the field
uniformly, so that they are combined into a new effective spin (or a
cluster). Then we effectively eliminate one spin degree of freedom,
and the local field at the new effective spin is $h'_i = h_i
h_j/J_{ij}$. These two decimation procedures are illustrated in
Fig.~\ref{F.RG}. Numerically the RG is processed until only one
cluster is left in the system. The ground state then consists of
independent  clusters each of which is frozen into a \emph{GHZ}
state:
$|C(n)\rangle = \frac{1}{\sqrt{2}}\left(|\uparrow\rangle^{\bigotimes
n} + |\downarrow\rangle^{\bigotimes n}\right).$
Since each \emph{GHZ} state will contribute either $1$  (we take
logarithms in base two) to the entropy if it consists of degrees of
freedom in both subsystems $A$ and $B$, or $0$  otherwise,
calculating the entanglement entropy between two subsystems is
reduced to a pure \emph{cluster counting} problem: $S(L)$ is
proportional to the number of clusters $N(L)$ that cross the
boundary between the two subsystems. Finally $S(L)$ is averaged over
different disorder configurations. In practice, $10^5-10^6$
configurations are used.

\begin{figure}[h]
\begin{center}
\includegraphics[
bbllx=30pt,bblly=30pt,bburx=849pt,bbury=624pt,%
     width=85mm,angle=0]{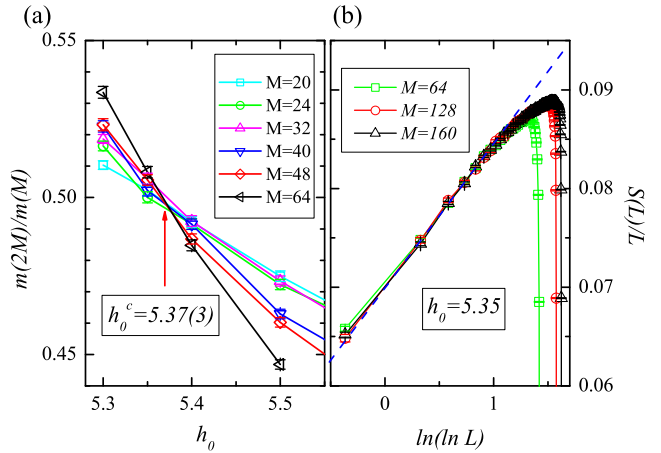}
\caption{(color online) (a): Finite-size scaling of magnetization
ratio given in Eq.~\ref{E.MagRatio}; (b): entropy per surface
$S(L)/L$ v.s. $\ln\ln L$ at critical field $h^c_0=5.35$, The dashed
line is a linear fit in $\ln\ln L$ scale.} \label{F.EntropyQCP}
\end{center}
\end{figure}

The foregoing technique has been applied with success in the  1D
case. The technical difference in 2D is first that the quantum
critical point is not exactly known. To locate it accurately we
study the scaling behavior of the average magnetization
$m(M)$~\cite{Karevski01}. At the critical field $h^c_0$,
\begin{equation}
\left.\frac{m(2M)}{m(M)}\right|_{h_0=h^c_0} =
2^{-x_m},\label{E.MagRatio}
\end{equation}
is independent of $M$, where $x_m$ is the anomalous dimension of the
bulk magnetization. Our result is given in
Fig.~\ref{F.EntropyQCP}(a). The critical field is estimated to be
$h^c_0 = 5.37\pm0.03$, with $x_m = 1.01\pm0.05$, which is consistent
with a previous RG study~\cite{Linetal00}. The entropy $S(L)$ is
calculated for various values of $h_0$. For both critical and
noncritical $h_0$, we find that the area law holds: $S(L)\sim L$ in
the leading term. The result of $S(L)/L$ for different system sizes
at critical $h^c_0 = 5.35$ is shown in Fig.~\ref{F.EntropyQCP}(b).

This conclusion is quite different from the one in a recent
study~\cite{Lin07}, where  a double-logarithmic modification of
 the area law  in the same  model was reported at the critical point. We
find
 that for small systems the  double-logarithmic fit is reasonable,
 but that for system size $M\geqslant128$, $S(L)/L$ increases
definitely slower than $\ln\ln L$ for $L\gtrsim 48$, strongly
suggesting $S(L)\sim L$ in the limit of $L\rightarrow\infty$,
without modification, and that the observation of Ref.~\cite{Lin07}
is biased by finite size effects. We also note that our results are
largely independent of the distribution of couplings, confirming the
idea of universal behavior for $S(L)$.

\begin{figure}[h]
\begin{center}
\includegraphics[
bbllx=30pt,bblly=30pt,bburx=859pt,bbury=643pt,%
     width=85mm,angle=0]{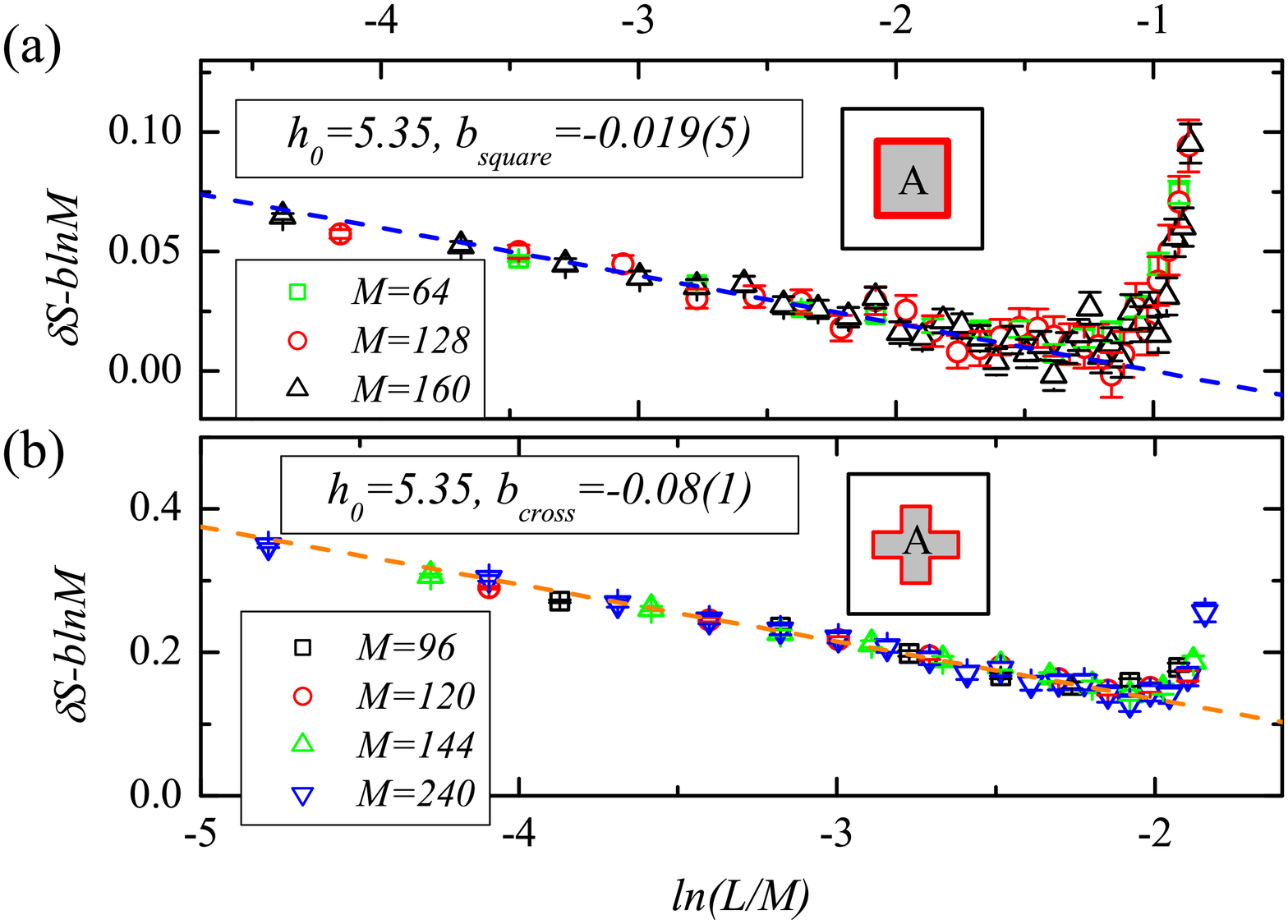}
\caption{(color online) (a): Scaling plots of $\delta S(L)$ to
reveal the subleading term of $S(L)$ at critical field $h^c_0=5.35$
for different geometries of subsystem $A$: a square (in (a)), and a
cross shape (in (b)). Dashed lines are linear fits in $\ln L$ scale
with $b_{square}=-0.019$ and $b_{cross}=-0.08$, respectively. }
\label{F.EntropyGeo}
\end{center}
\end{figure}

Having established the validity of the area law in this system, it
is natural to investigate subleading terms.  We thus consider
$\delta S(L) \equiv  2S(L) -S(2L)$, in which the terms linear in $L$
cancel exactly. We find in both disordered and ordered phases that
$\delta S(L)$  saturates to a constant term, indicating $S(L) =
aL + c$. Meanwhile, at the  critical point, we find that $\delta
S(L)$ scales linearly as $\ln L$, suggesting
\begin{equation}
S(L) = aL + b\ln L +c,
\end{equation}
i.e., a logarithmic correction to the area law. The coefficient of
this logarithmic correction is determined to be $b=-0.019\pm0.005$
through finite-size scaling in Fig.~\ref{F.EntropyGeo}(a). To our
knowledge, this is the first instance of such behavior in disordered
2D systems, and the second instance in all 2D systems after the
examples in Ref.~\cite{FradkinMoore06} for a class of conformal
quantum critical models with dynamical exponent $z=2$. There is no
reason to expect that the $\ln L$ term we find in the critical RTFI
model has much to do with the latter. This can be substantiated by
calculating the amplitudes of the logarithmic term for different
geometries, which obey some precise relations in the case of
Ref.~\cite{FradkinMoore06}. As an example, we considered a cross
shape geometry as shown in Fig.~\ref{F.EntropyGeo}(b). In this case
as well, we can resolve a $\ln L$ term in $S(L)$ in addition to the
area law contribution, with the coefficient $b_{cross} =
-0.08\pm0.01$. We can then calculate and compare the ratios  in our
model, where we obtain $b_{cross}/b_{square} \thickapprox 4$, and in
the conformal quantum critical models where $b_{cross}/b_{square} =
3$ exactly. This implies that the $\ln L$ term in $S(L)$ in our
model most probably has  a different origin, a non surprising
conclusion since, for the  IRFP, $z\rightarrow\infty$.

\begin{figure}[h]
\begin{center}
\includegraphics[
bbllx=40pt,bblly=40pt,bburx=884pt,bbury=586pt,%
     width=85mm,angle=0]{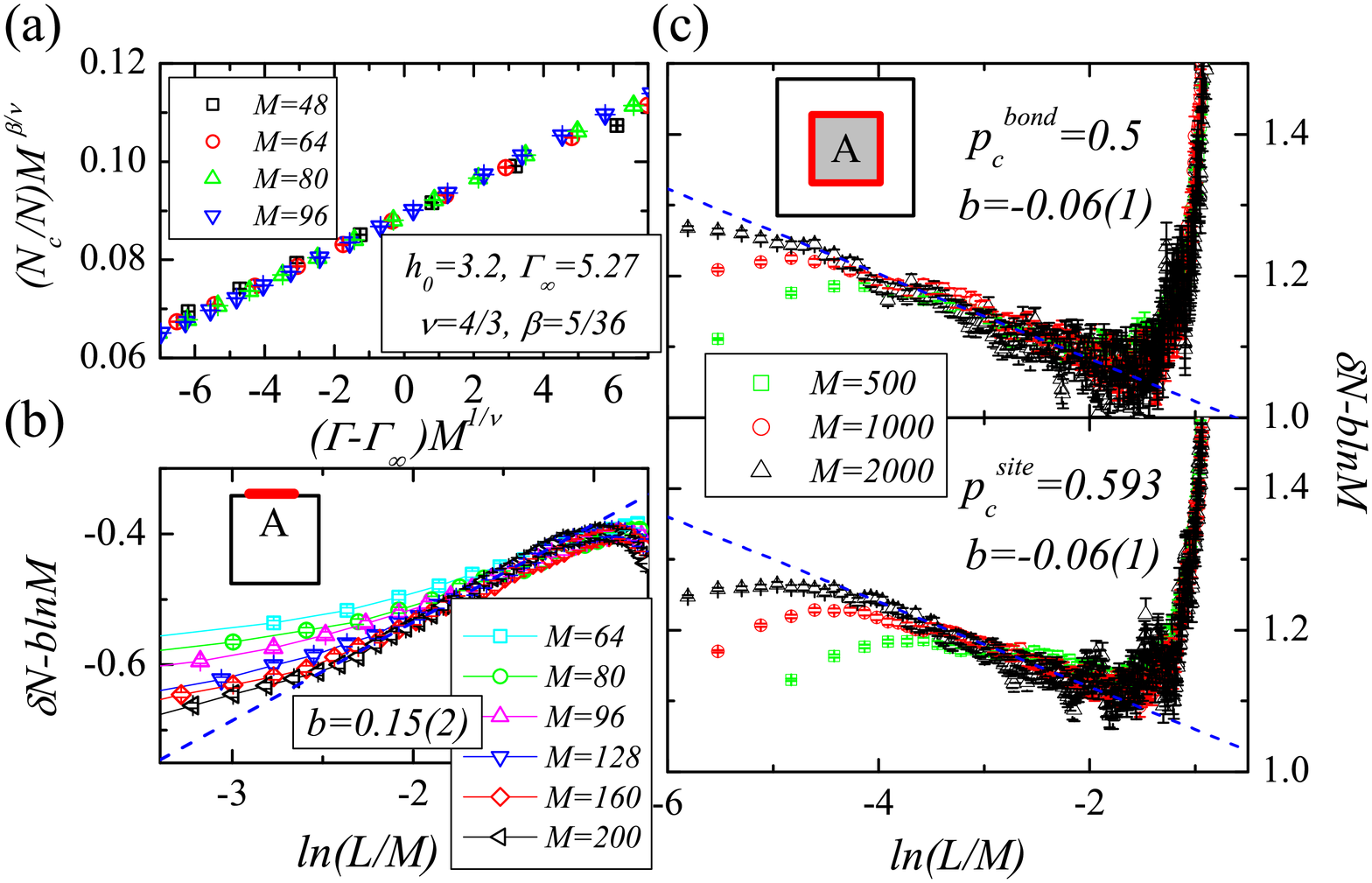}
\caption{(color online) (a): Scaling of the largest active cluster
size $N_c$ during RG shows a signature of classical percolation at a
finite energy scale; (b): scaling of $\delta N$ at percolation
threshold $\Gamma_\infty=5.27$ for $h_0=3.2$, $A$ is a line interval
on the boundary; (c): scaling of $\delta N$ at bond (upper) and site
(lower) classical percolation thresholds, $A$ takes the geometry of
a square. In (b) and (c), dashed lines are linear fits in $\ln L$
scale with corresponding slope $b$ marked on the plot.}
\label{F.NClusterPerc}
\end{center}
\end{figure}

To better understand the $\ln L$ term in $S(L)$ at the IRFP in 2D
RTFI model, we notice that there is a striking difference between
the model in 2D and in 1D. In 2D for any $h_0<h^c_0$ there is a
finite-temperature phase transition at
$T_c(h_0)$~\cite{2DRandomIsing}: the 
IRFP in 2D can then be considered as an extension of this
finite-temperature transition right down to  $T=0$. Through the
SDRG, the transition to a ferromagnetically ordered phase can be
mapped to a percolation transition in 2D~\cite{Motrunich00}: the
magnetic transition corresponds to the  development of an infinite
percolating spin cluster during RG. It is widely expected that this
percolation process at the IRFP (which occurs at energy scale
$\Omega_{\infty}=0$) is different from the one at finite
temperature, since at $h_0=h^c_0$ the critical behavior is
controlled by quantum fluctuations. This leads one to think of the
IRFP as a type of ``quantum percolation'', with fractal dimension
$d_f=2-x_m\approx1.0$. For $h_0<h^c_0$ meanwhile,
the percolation takes place at finite energy scale
$\Omega_\infty\sim T_c$ in the RG, and is expected to be in the
universality class of conventional classical percolation~\cite{Motrunich00}. 
We have confirmed the classical percolation picture at finite energy
scale by studying the scaling of largest active cluster size during
RG. Some  numerical results at $h_0=3.2$ are presented in
Fig.~\ref{F.NClusterPerc}(a). The percolation threshold is at
$\Gamma_\infty=\ln(\Omega_0/\Omega_\infty)=5.27\pm0.02$, where the
extracted exponents $\beta$ and $\nu$ take the values of classical
percolation indeed.

The number of clusters crossing the boundary between two subsystems
can be investigated at these percolation transitions as well (even
though it does not correspond to an entanglement entropy except when
$h_{0}=h_{0}^{c}$). It is easy to see then that a $\ln L$ correction
to the area law is expected and related to conformal invariance,
even though it depends on more complicated parameters that the
central charge and the topology (in contrast with the example in
Ref.~\cite{FradkinMoore06}).

To see this, consider the ``baby'' case where  the subsystem $A$
takes the geometry of a line interval of length $L$ on the boundary
of the lattice, so the boundary between the two subsystems is $A$
itself. Now  $N(L)$  is simply equal to the number of clusters
touching $A$. Its scaling can be studied using CFT techniques. For
this, consider first the problem on the upper half complex plane,
with $A$ on the real axis. Use the well known expansion of the
partition function of the critical $Q$-state Potts model in terms of
clusters, or equivalently, dense loops~\cite{Nienhuis}:
\begin{equation}
Z = \sum_P \sqrt{Q}^{N_P}
\end{equation}
where every loop gets the same weight $\sqrt{Q}$. Now introduce a
boundary conditions changing operator (BCCO)~\cite{Cardy89} $\phi_y$
such that the two point function of $\phi$ is defined through a sum
similar to the one for $Z$, but loops touching the boundary between
the two insertions $\zeta_1$ and $\zeta_2$ get a different weight
$y$ instead of $\sqrt{Q}$:
\begin{equation}
<\phi_y(\zeta_1)\phi_y(\zeta_2)> = \frac{1}{Z} \sum
\sqrt{Q}^{N_P} \left({y\over \sqrt{Q}}\right)^{N^b_P(\zeta_1,\zeta_2)},
\end{equation}
where $N^b_P(\zeta_1,\zeta_2)$ gives the number of loops touching
the boundary between $\zeta_1$ and $\zeta_2$ located on the real
axis. We expect the two point function to have the following scaling
form:
\begin{equation}\label{E.scaling}
<\phi_y(\zeta_1)\phi_y(\zeta_2)> \sim e^{-f(y)|\zeta_1-\zeta_2|}
|\zeta_1-\zeta_2|^{-2h(y)},
\end{equation}
where $f(y)$ is the boundary free energy induced by the modified
weight on the boundary, and the exponent $h(y)$ is the anomalous
dimension of the BCCO. Now differentiate the two point function of
BCCOs with respect to the weight $y$, then take the limit
$y=\sqrt{Q}$. This leads to
\begin{equation}
N^b_P(L) = aL + b\ln L,
\end{equation}
where
\begin{eqnarray}
b &=& -2\sqrt{Q}\left.\frac{\partial h(y)}{\partial
y}\right|_{y=\sqrt{Q}},
\end{eqnarray}
and $L=|\zeta_1-\zeta_2|$.

With the exact expression of $h(y)$~\cite{JacobsenSaleur06} we
obtain
\begin{equation}
b = \frac{1}{2\pi p} \sqrt{Q(4-Q)}. \label{E.prefactor}
\end{equation}
For percolation ($Q=1$),
$b_{perc}=\frac{\sqrt{3}}{4\pi}\thickapprox 0.1378$,
reproduces an early result by Cardy~\cite{Cardy01}. 
 But in Eq.~\ref{E.prefactor} we generalize Cardy's result to
general $Q$, and it is interesting to see that the $\ln L$ term
vanishes at $Q=0$ and $Q=4$. It is also remarkable to see that $b$
is related to the derivative of the anomalous dimension of BCCO, but
not the central charge. This $\ln L$ term is also observed in our RG
calculation at finite $\Gamma_\infty$. In
Fig.~\ref{F.NClusterPerc}(b) we show the scaling of $\delta N \equiv
2N(L) - N(2L)$ at $\Gamma_\infty = 5.27$ for $h_0=3.2$. $b$ is
estimated to be $0.15\pm0.02$, in agreement with the analytical
result. This further confirms that  the universality class at finite
$\Gamma_\infty$ is classical percolation. Interestingly,  we find
numerically for this case  that $|b|<0.01$ at the quantum critical point,
consistent again with the idea of a different universality class when
$\Gamma_\infty\rightarrow\infty$.

Going back to the original problem  where  the subsystem $A$ takes
the geometry of a $L\times L$ square, we have not derived a similar
 analytical result for ordinary percolation. But $N(L)$ can of course
 be calculated numerically. To get better
scaling for large systems,
we turn to a direct study of percolation. In
Fig.~\ref{F.NClusterPerc}(c), $\delta N 
$ data
at percolation threshold are shown. As well expected, $\delta N \sim
\ln L$ is resolved for both bond and site percolation, and the
coefficient of the $\ln L$ term takes the same value $b =
-0.06\pm0.01$, in agreement with the idea that this term is
universal. Note that we observe a negative value of $b$ for
subsystem $A$ a  square, just as in the case of the RTFI model.
This is opposite to the sign of $b$ in classical percolation when
$A$ is an interval.

The observation of a $\ln L$ term in $N(L)$ in percolation  makes
the presence of a similar term at the IRFP most likely:
 there will in fact always be such a term at energy scale
 $\Omega_{\infty}$. When this scale
is finite, the coefficient $b$ takes the value of
classical percolation, $b=-0.06\pm0.01$. But when
$\Omega_\infty\rightarrow0$, i.e., at the quantum critical point,
quantum fluctuations become dominant, leading to a quantum
percolation belonging to different universality class. A different
$b$ value, $b=-0.019\pm0.005$, reflecting this difference is then
observed.


In summary, we have calculated the entanglement entropy of a 2D RTFI
model by using a numerical SDRG method. In contrast to what is
claimed in a recent preprint, we find that the leading term of the
entropy follows the \emph{area law} and depends linearly on the
block size $L$ in both critical and non-critical phases. However, a
$\ln L$ correction to the area law is discovered at criticality.
While the presence of this correction may not have been expected
from the entanglement point of view, it is very natural once the
problem is reformulated geometrically. Indeed, the problem of
counting clusters touching a boundary in 2D classical percolation is
easily argued to give rise to sub logarithmic corrections, while the
entanglement entropy in the RTFI model at criticality can be
reformulated as a similar problem but in a different,
 ``quantum percolation'' universality class.

Useful discussions with N. Bray-Ali, L. Ding, J. Latorre, W. Li, J.
E. Moore, and G. Refael are gratefully acknowledged. R. Y.
acknowledge hospitality at Service de Physique Th\'{e}orique, CEN
Saclay. This work was supported by DOE, Grant No. DE-FG02-05ER46240.
Computational facilities have been generously provided by the
HPCC-USC Center.


\begin{thebibliography}{99}
\bibitem{footnote} A question of vocabulary: such subleading terms (corrections) should
not be confused with terms that modify the leading behavior, which
we call here modifications.
\end{thebibliography}
\end{document}